\documentclass[12pt,letter,doublespace]{article}
\usepackage{amssymb}
\usepackage{amsmath} 
\usepackage{graphics}
\usepackage{epsfig}
\usepackage{graphicx, subfigure}
\usepackage{cite}
\usepackage{amsthm}
\usepackage{color}
\usepackage{booktabs}
\usepackage{bbm}
\usepackage{caption}

\usepackage{harvard}

\usepackage{floatrow}
\newfloatcommand{capbtabbox}{table}[][\FBwidth]
\usepackage{blindtext}

\newcommand{\vx}{{\bf x}}
\newcommand{\vy}{{\bf y}}
\newcommand{\vX}{{\bf X}}

\newcommand{\vY}{{\bf Y}}

\newcommand{\vmu}{\mbox{\boldmath $\mu$}}

\newcommand{\bqa}{\begin{eqnarray*}}
\newcommand{\eqa}{\end{eqnarray*}}
\newcommand{\bqan}{\begin{eqnarray}}
\newcommand{\eqan}{\end{eqnarray}}
\newcommand{\bit}{\begin{itemize}}
\newcommand{\eit}{\end{itemize}}
\newcommand{\ben}{\begin{enumerate}}
\newcommand{\een}{\end{enumerate}}
\newcommand{\beq}{\begin{equation}}
\newcommand{\eeq}{\end{equation}}
\newcommand{\bdes}{\begin{description}}
\newcommand{\edes}{\end{description}}

\begin{document}
\bibliographystyle{dcu} 
\citationstyle{dcu}

\title{\large {\bf Genetic Algorithm for Constrained Optimization with Stochastic Feasibility Region with Application to Vehicle Path Planning} }
\author{\sc Adriano Z. Zambom, Julian A. A. Collazos
 and Ronaldo Dias\\
Universidade Estadual de Campinas - UNICAMP}
\date{}
\maketitle{}

\section*{Abstract}
In real-time trajectory planning for unmanned vehicles, on-board sensors, radars and other instruments are used to collect information on possible obstacles to be avoided and pathways to be followed. Since, in practice, observations of the sensors have measurement errors, the stochasticity of the data has to be incorporated into the models. In this paper, we consider using a genetic algorithm for the constrained optimization problem of finding the trajectory with minimum length between two locations, avoiding the obstacles on the way. 
To incorporate the variability of the sensor readings, we propose a more general framework, where
the feasible regions of the genetic algorithm are stochastic. In this way, the probability that a possible solution of the search space, say $x$, is feasible can be derived from the random observations of obstacles and pathways, creating a real-time data learning algorithm. 
By building a confidence region from the observed data such that its border intersects with the solution point $x$, the level of the confidence region defines the probability that $x$ is feasible. We propose using a smooth penalty function based on the Gaussian distribution, facilitating the borders of the feasible regions to be reached by the algorithm. 


\textbf{Keywords:} constrained optimization; stochastic feasible regions; penalty function; autonomous vehicle; nonparametric curve estimation.

\pagestyle{plain}
\setcounter{page}{1}
\setlength{\textheight}{9.0in}
\setlength{\topmargin}{-0.5in}

\section{Introduction}

\indent \indent Genetic Algorithms are very powerful and efficient search metaheuristics inspired by natural selection and concepts of evolution. Dating back to \citeasnoun{Holland1978} and \citeasnoun{Goldberg1989}, this methodology has become very popular in several fields such as engineering, bioinformatics and basically any application of nonlinear optimization. However, the simple strategies of genetic algorithms completely disregard possible constraints, and hence candidate solutions of the optimization problem found without any control of these constraints may not be feasible. Hence, it is necessary to consider adaptations of the algorithms in order to solve constrained optimization problems.

Researchers have recently developed several procedures and techniques to handle constrained optimization, not allowing candidate solutions to stay outside the feasibility region. An interesting and throughly review can be found in \citeasnoun{Coello2002}. The most common way of incorporating constraints into a genetic algorithm is to add a penalty function to the minimization(or maximization) problem, so that infeasible solutions are penalized according to the degree of violation. By doing this, the constrained optimization problem becomes a nonlinear unconstrained optimization problem. 

When the penalty functions are not well chosen, they may become a burden to the algorithm extending its time to converge. If the penalty is too low to the point that it is negligible with respect to the objective function, then the algorithm will have a longer expected time to converge, for it will spend too much time searching infeasible regions. On the other hand, if the penalty is too high and the optimum lies near the boundary of feasibility, the algorithm will push the candidates inside the feasible region, and will have problems reaching the optimum. In general, genetic algorithms should follow the minimum penalty rule (see \citeasnoun{Davis1987}, \citeasnoun{RicheHaftka1993}, \citeasnoun{SmithTate1993}), keeping the penalty as low as possible. 

Inspired by modern unmanned vehicle problems, we propose using a genetic algorithm to search for the trajectory with minimum length between two locations. In most applications of autonomous vehicles, on-board sensors and radars are used to collect information on possible obstacles to be avoided and pathways to be followed. Thus, genetic algorithms are appropriate methods to deal with the several local minima in the trajectory search. Moreover, in practice the radars of the vehicles have measurement errors, so that a stochastic approach has to be incorporated when computing the feasible regions.
In this paper we address the general problem that the feasibility region of solutions in a genetic algorithm is not crisply known, but instead, its stochasticity can be determined by the observation of a random process, enabling the estimation of the probability that a possible solution of the algorithm is feasible. In addition, we propose a different definition for the feasibility region, facilitating the handling of disjoint regions with penalty functions. Finally, we propose a new penalty function, whose smoothness and continuity make it suitable for the stochastic feasibility regions, and help deal with the border problem of feasibility regions in genetic algorithms.


The remainder of the paper is as follows: In Section \ref{sec.union}, we propose a new definition for the unconstrained optimization problem for feasibility regions that are union of disjoint sets. Section \ref{sec.stoc.feas} contains the methodology proposed for when the feasible regions are stochastic, originated from observations of random processes. In Section \ref{sec.experimental} we run simulations on experimental examples, and in Section \ref{sec.robot} we present an application of the methodology for autonomous vehicle path planning.

\section{Feasibility Regions with Union of Sets} \label{sec.union}
\indent \indent The problem of constrained optimization is usually written as a nonlinear optimization problem defined by a set of linear and nonlinear constraints, specifically
\bqan\label{eq.minimize}
&&\mbox{minimize  } f(\vx) \hspace{1cm} \vx \in F \subseteq S \subseteq \mathbb{R}^n\nonumber \\
&&\mbox{subject to} \nonumber \\
&& g_i(\vx) \leq 0 \hspace{1cm} i = 1, \ldots, p \nonumber \\
&& h_j(\vx) = 0 \hspace{1cm} j = 1, \ldots, q,
\eqan
where $\vx$ is a vector of possible solutions, $F$ is the feasibility set and $S$ is the search space. Note that there are $p$ nonlinear constraints and $q$ linear constraints which basically define the feasibility set $F$. This definition is widely accepted and includes most types of feasibility regions, as long as the functions $g_i, i = 1, \ldots, p$ and $h_j, j = 1, \ldots, q$ are carefully chosen according to the problem at hand.

There are several ways of dealing with constraints in genetic algorithms, including repairing unfeasible individuals (\citeasnoun{MichalewiczNazhiyath1995}, \citeasnoun{ChootinanChen2006}), preserving feasibility of solutions (\citeasnoun{MichalewiczJanikow1993}, \citeasnoun{KozielMichalewicz1999}), rankings (\citeasnoun{RunarssonYao2000}), just to cite a few. However, the most accepted approach is to make use of penalty functions. Penalized constrained optimization has received extensive attention in recent years, see for example \citeasnoun{LinWu2004}, \citeasnoun{WangEtAl2009}, \citeasnoun{MontemurroEtAl2013} and references there in.
The general formulation of the problem using the penalization approach is to minimize the function
\bqan \label{eq.penal} 
&& \ell(\vx) = f(\vx) + p(\vx), \mbox{  where  } \nonumber\\
&& p(\vx) = \sum_{i=1}^p r_i \max(0,g_i(\vx))^\alpha + \sum_{j=1}^q c_i |h_j(\vx)|^\beta,
\eqan
$r_i$ and $c_j$ are penalty parameters, $\alpha$ and $\beta$ are constants (usually equal to 1 or 2).

Let us now concentrate for a moment on the restrictions implied by the inequalities $g_i(\vx) \leq 0, i = 1, \ldots, p$. Even though the definition of the problem in (\ref{eq.minimize}) can be used for most situations, it is really suitable for feasibility regions that are obtained by the intersection of the $p$ regions defined by the inequalities. If the feasible set $F$ is composed by the union of several  possibly disjoint regions, it may not be easy to find inequalities that combined would yield $F$. Explicit work on disjoint regions is not very common,  and in most cases it is only considered as a direct application without much formal strategies, see for instance \citeasnoun{Liu2010}, \citeasnoun{Jabr2012}.

To facilitate dealing with difficult feasibility regions, which are composed of union of several sets, we propose handling the constrained optimization problem with another formulation. Suppose that $F$ can be written as the union of $m$ feasible regions $F_k, k = 1, \ldots, m$, which are defined by a set of $p_k, k = 1, \ldots, m$ inequalities respectively. Note that by defining $F = F_1\cup\ldots \cup F_m$, a solution $\vx$ which belongs to one of the sets $F_k$ but does not belong to any other $F_{k'}, k \neq k'$, should not receive any penalty. Hence, the simple somation of penalties for each of the inequalities may penalize feasible solutions. Formally, we propose the following model. Suppose that we want to minimize $f(\vx)$ subject to
\bqan\label{eq.nonlinear_constr}
&& g_1^{(1)}(\vx) \leq 0, \ldots, g_{p_1}^{(1)}(\vx) \leq 0 \nonumber\\
&& \mbox{or} \nonumber\\
&& g_1^{(2)}(\vx) \leq 0, \ldots, g_{p_2}^{(2)}(\vx) \leq 0 \nonumber\\
&& \mbox{or} \nonumber\\
&& \vdots \nonumber\\
&& g_1^{(m)}(\vx) \leq 0, \ldots, g_{p_m}^{(m)}(\vx) \leq 0
\eqan
where each of the above set of inequalities define a set/region of feasibility $F_k = \{\vx: g_1^{(k)}(\vx) \leq 0, \ldots, g^{(k)}_{p_k}(\vx) \leq 0\}, k = 1,\ldots,m$, so that the feasible region is $F = \cup_{k = 1}^m F_k$. In this case, we propose using the following penalty function
\bqa
p(\vx) = \min_{k = 1,\ldots,m}\left(\sum_{i = 1}^{p_k}r_i^{(k)} \max(0,g_i^{(k)}(\vx))^{\alpha_k}\right).
\eqa
It is easy to see that if any of the $m$ sets of constraints is satisfied, the penalty is equal to 0. Another penalty that would be suitable for this situation is
\bqa
p(\vx) = \left[\sum_{k = 1}^{m}\sum_{i = 1}^{p_k}r_i^{(k)} \max(0,g_i^{(k)}(\vx))^{\alpha_k}\right]\mathbbm{1}\left(\sum_{i = 1}^{p_k} \max(0,g_i^{(k)}(\vx))^{\alpha_k} > 0, k = 1,\ldots,m\right),
\eqa
where $\mathbbm{1}$ is the indicator function
\bqa
\mathbbm{1}(A) = \begin{cases}
1 & \mbox{  if condition   } A \mbox{  is satisfied}\\
0 & \mbox{ otherwise}.
\end{cases}
\eqa
Note that it is enough to have one of the $m$ sets of $p_k$ constraints satisfied in order for the penalty to be 0 since the indicator function will be 0.

The same approach can be used for the linear constraints, defining
\bqan\label{eq.linear_constr}
&& h_1^{(1)}(\vx) = 0, \ldots, h_{q_1}^{(1)}(\vx) = 0 \nonumber\\
&& \mbox{or} \nonumber\\
&& h_1^{(2)}(\vx) = 0, \ldots, h_{q_2}^{(2)}(\vx) = 0 \nonumber\\
&& \mbox{or} \nonumber\\
&& \vdots \nonumber\\
&& h_1^{(s)}(\vx) = 0, \ldots, h_{q_s}^{(s)}(\vx) = 0.
\eqan
Thus, for the optimization problem with the inequalities constraints in (\ref{eq.nonlinear_constr}) and linear constraints in (\ref{eq.linear_constr}), we define the following penalty
\bqa
p(\vx) = \min_{k = 1,\ldots,m}\left(\sum_{i = 1}^{p_k}r_i^{(k)} \max(0,g_i^{(k)}(\vx))^{\alpha_k}\right) + \min_{k = 1,\ldots,s}\left(\sum_{i = 1}^{q_k}c_i^{(k)}|h_i^{(k)}(\vx)|^{\beta_k}\right).
\eqa



\section{Stochastic Feasibility} \label{sec.stoc.feas}

\indent \indent In the last decades, an enormous amount of work has been done on constrained optimization using genetic algorithms. Several types of penalty functions have been proposed to reduce the fitness of infeasible solutions according to the degree of constraint violation. When defining a penalty function, it is sightlessly assumed that the feasibility region $F$ is fully known (crisp set). This means that all the constraints (e.g. (\ref{eq.minimize})) that form $F$ are  hypothesized to be known a priori.

Consider the situation where the feasible set is not well defined or entirely known in advance, but instead it is possible to derive the probability that a solution of the search space $\vx$ is feasible. This set up has a wide range of applications in diverse fields such as engineering, mathematics, operations research, water management, optimization of chemical processes and many others, see for example \citeasnoun{DupacovaEtAl1991},
\citeasnoun{HenrionEtAl2001} and \citeasnoun{HenrionMoller2003}.

Assume we have the formulation of the optimization problem as described in Section \ref{sec.union} such that the feasibility region is $F = F_1\cup\ldots\cup F_m$. Also, assume that we observe the random vectors
\bqa
&& \vX^1_1, \ldots, \vX^1_{n_1}\nonumber\\
&& \vX^2_1, \ldots, \vX^2_{n_2}\nonumber\\
&& \vdots\nonumber\\
&& \vX^m_1, \ldots, \vX^m_{n_m}\nonumber
\eqa
where $\vX^k_1, \ldots, \vX^k_{n_k}, k = 1,\ldots,m$ are i.i.d. independent of $\vX^{k'}_1, \ldots, \vX^{k'}_{n_1}, k' \neq k$. The main idea is that, for a possible solution $\vx$ of the genetic algorithm, the observations $\vX^k := (\vX^k_1, \ldots, \vX^k_{n_k})$ yield the probability $\gamma^k_{n_k}(\vx)$ that $\vx$ is in the $k$-th feasibility region, i.e., 
\bqa
P(\vx \in G_{\gamma_{n_k}^k}^k(\vX^k)) = \gamma_{n_k}^k(\vx),
\eqa
where $G_{\gamma_{n_k}^k}^k(\vX^k)$ is the confidence region of feasibility for the set $F_k$ estimated from the random vectors $\vX_1^k, \ldots,\vX_{n_k}^k$.
 We will omit the subscript $n_k$ of $\gamma^k(\vx)$ for ease of notation. 
 By building the confidence region such that its border intersects with the solution point $\vx$, the level of confidence of this region defines the probability $\gamma^k(\vx)$ that the solution $\vx$ is feasible. The calculation of $\gamma^k(\vx)$ clearly reminds that of the p-value, for the probability is computed using the tail of the distribution.
 
By construction, the sizes of confidence regions depend on the standard deviation estimated from the data. As the number of observations $n_k$ grows, the volume of the confidence region tends to be smaller, approaching the true location of the feasibility region. In this way, the solution of the genetic algorithm that runs with the stochastic feasible sets gets closer to the solution of a genetic algorithm with the true feasible region whenever more observations are made. This property can be used to build an iterative algorithm, where the feasible regions are constantly updated with new observations of the random process, creating a real time data learning algorithm. Such approach can be improved by using a Markovian structure to estimate the probabilities, as in \citeasnoun{DiasEtAl2012}, however further exploration of this procedure is needed and will be left for future investigation. 

One may find the definition and construction of the proposed methodology similar to that of fuzzy theory for chance constrained programming. Since \citeasnoun{CharnesCooper1959} and \citeasnoun{MillerWagner1965}, several authors have considered the problem of chance constrained programming throughout the years, see for example \citeasnoun{CooperEtAl1998}, 
\citeasnoun{RachevRomisch2002},
\citeasnoun{Calafiore2006},
\citeasnoun{ChenEtAl2010},
 \citeasnoun{Kuecuekyavuz2012}, and references therein.
 It consists basically of minimizing $f(\vx)$ subject to $P(\xi | \ell_k(\vx,\xi) \leq 0, k = 1,\ldots,m) \geq \tilde{\alpha}$, where $\xi$ is a random variable, $g_k(\vx,\xi)$ are stochastic constraint functions and $\tilde{\alpha}$ is the choice of level. In this definition, a solution $\vx$ that satisfies this probability constraint is called feasible. 
 
The methodology proposed in this paper differs conceptually from that of chance constrained programming in the sense that, the stochastic constraints that compose the feasibility regions are fundamentally based on data learning. The proposed approach states that, given a possible solution $\vx$, one can build a confidence interval/region that intersects with $\vx$ such that the level of confidence of this region will be taken as the probability of feasibility estimated from the data. Nothing is specified about the form of the constraints or how the random process (data) will have influence on it. Thus, this stochastic approach is very general, to such a degree that it can accommodate any type of error in the feasibility regions, more specifically, the definition of how the random vectors $\vX$ will be used to build the confidence region $G^k$ is completely free and problem specific. Chance constrained programming on the other hand, specifies a priori a deterministic structure that is taken to be randomly perturbed, without learning from data.

In the usual formulation, where the feasibility region $F$ is crisp, penalty functions are equal to 0 when no violation occurs, and increase with the degree of constraint violation, having as consequence discontinuities on the borders, jumping from 0 to $r_i$ (see Equation (\ref{eq.penal})). Due to the fact that probabilities are always between 0 and 1, we propose the following smooth penalty function
\bqan \label{penalty}
p(\vx) = \min_{k = 1\ldots m} \Psi\Phi(Z_\alpha+\sqrt{H}(\tilde{\alpha} - \gamma^k(\vx))),
\eqan
where $\Psi, \alpha$ and $H$ are running parameters, $\tilde{\alpha}$ represents the significance level (user choice) and $\Phi$ is the cumulative standard Normal distribution. A similar penalty function was used by \citeasnoun{DiasEtAl2010}.

 By using such penalty function, and choosing wisely the slope, we do not penalize too much individuals  of the genetic algorithm which are near the border of feasibility, allowing their information to be passed on to next generations. In this way, the minimum penalty rule is alleviated, and minimums near the border can be found faster.
Moreover, the proposed penalty function is not a hard thresholding type, instead it has a slope and a maximum that can be controlled according to the specific problem at hand. These are considered tuning parameters, and can be estimated with cross-validation or other adequate procedures.

In the next section we will present two different scenarios with simulations where this definition is suitable.

\section{Experimental Studies} \label{sec.experimental}

\subsection{Unknown Location of Feasible Circles}

\indent \indent Suppose we want to find the point $\vx = (x_1, x_2)$ that minimizes the two dimensional Rastrigin function 
\bqa
f(\vx) = 20 + \sum_{i=1}^{2}(x_i^2 - 10\cos(2\pi x_i)).
\eqa

\begin{figure}
\centering
\begin{minipage}{.5\textwidth}
  \centering
  \includegraphics[width=.7\linewidth]{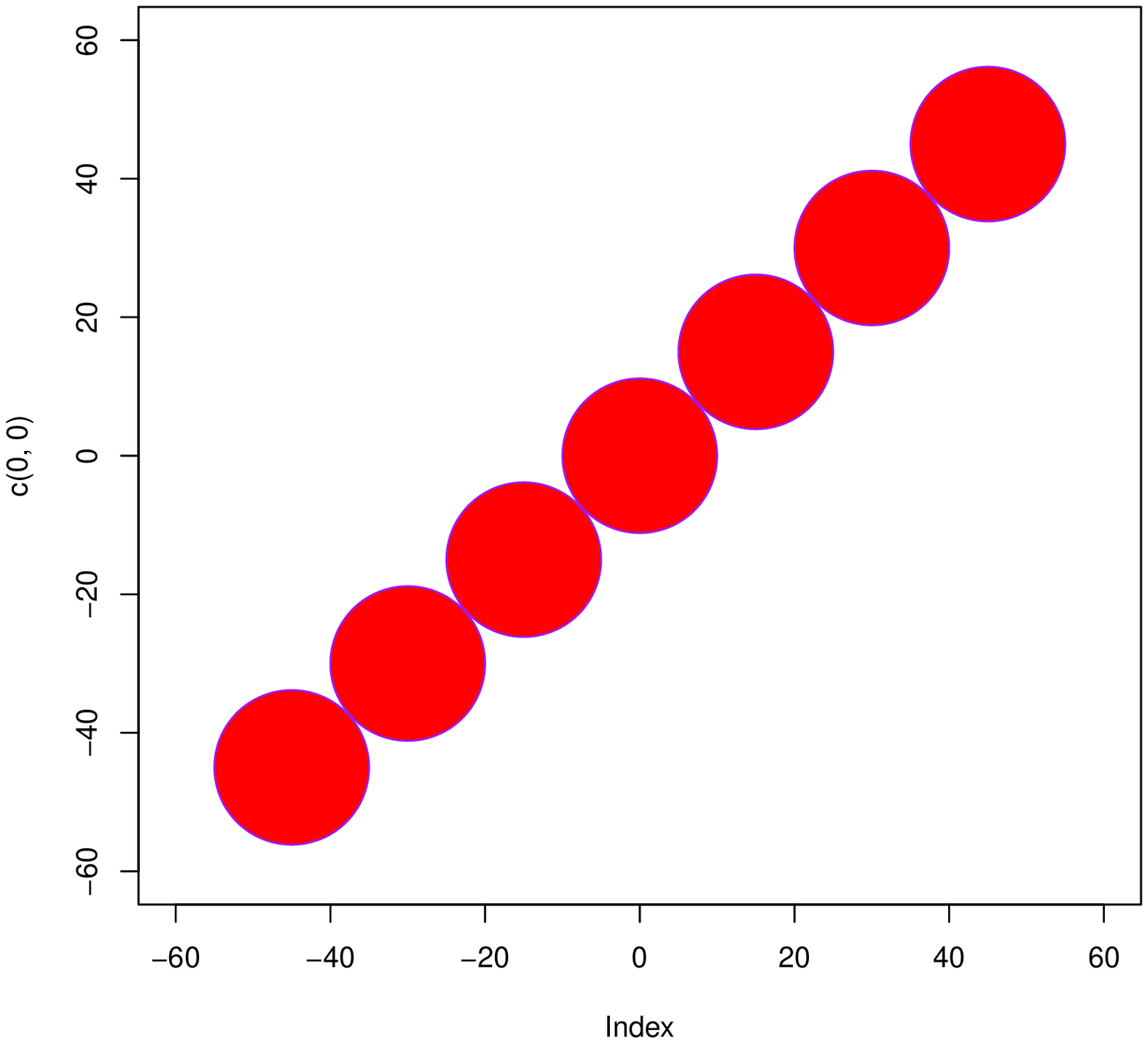}
  \caption{Left: Feasible region in red. Right: Example of observations of $\vX^1$ and $\vX^2$}{}
  \label{Fig.Rastrigin_F}
\end{minipage}%
\begin{minipage}{.5\textwidth}
  \centering
  \includegraphics[width=.7\linewidth]{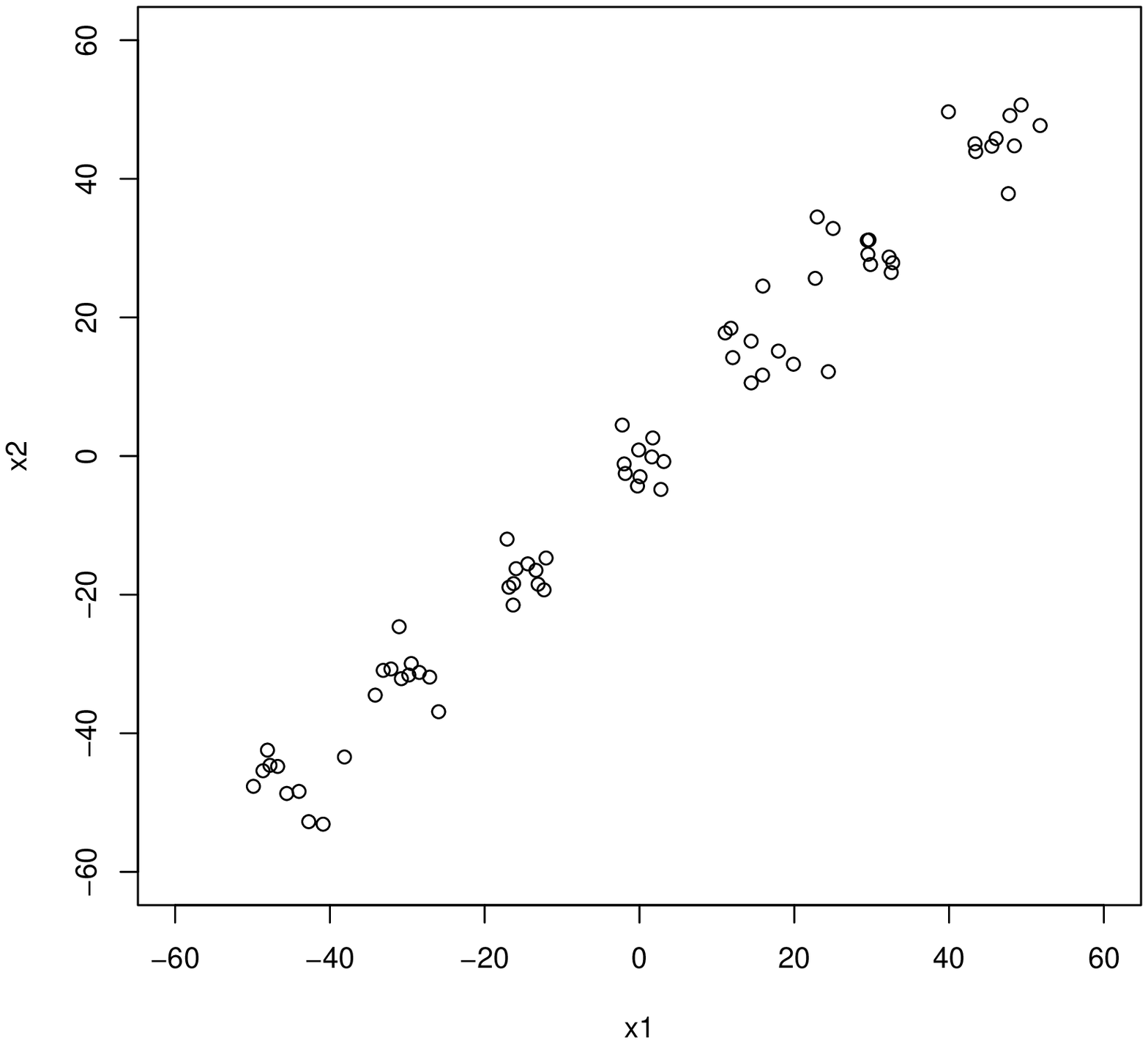}
\end{minipage}
\end{figure}

Assume that the search space is $S = \{\vx: -60 < x_1 < 60, -60 < x_2 < 60\}$ and the feasible set $F$ is given by 
\bqa
F = F_1\cup\ldots\cup F_m,
\eqa
where
\bqa
F_k= \{\vx: (x_1 - (15k - 60))^2 + (x_2 - (15k - 60))^2 - 10 \leq 0\},
\eqa
for $m = 7$. Basically, the feasible set $F$ is composed by the union of the circles in Figure \ref{Fig.Rastrigin_F}.

Consider the situation where the feasible set $F$ is not completely known, i.e., we can not verify exactly the feasibility conditions as stated in (\ref{eq.nonlinear_constr}). Suppose that it is known that the feasible regions $F_k, k = 1,\ldots, m$ are circles of known radius $r = 10$, but the location of each circle center is unknown. Then, given a set of $n$ observations of each of the random vectors $\vX^1,\ldots, \vX^m$, we can estimate a confidence region that covers the true center of the circle using the mean vector of the bivariate distributions. The confidence region that intersects with the solution point $\vx$ will be used to determine the probability that this point is feasible.

In this scenario, we simulate
\bqa
&&\vX^1_1, \ldots, \vX^1_n \sim N(\vmu_1,I_1)\\
&&\vdots\\
&&\vX^m_1, \ldots, \vX^m_n \sim N(\vmu_m,I_m),
\eqa
where
$\vmu_k = (15k - 60, 15k - 60)^T$ and 
$$I_k = 
\begin{pmatrix}
(10/3)^2 & 0\\
0 & (10/3)^2 
\end{pmatrix}, k = 1,\ldots,m.$$ Note that each $\vmu_k$, representing the mean of the random vector $\vX^k$, is exactly at the center of the feasible region $F_k$. 


In the Genetic Algorithm, given an individual $\vx$ of the population, define the probability that it is in the feasible set as:
\begin{enumerate}
\item if $d(\vx,\bar{\vX}^k) > r$, build a $100(1-\gamma^k(\vx))\%$ confidence ellipse $G^k(\gamma^k(\vx)): d(\vx, G^k(\gamma^k(\vx))) = r$, hence, the probability that $\vx$ is feasible for the $k$-th constraint is equal to $\gamma^k(\vx)$.
\item if $d(\vx,\bar{\vX}^k) \leq r$, then the probability that $\vx$ is feasible for the $k$-th constraint is approximately equal to $\gamma^k(\vx) \approx 1$,
\end{enumerate}
where $d(\vx,G^k(\gamma^k(\vx))) = \inf\{d(\vx, \vy): \vy \in G^k(\gamma^k(\vx))\}$ for a distance $d(.,.)$ (Euclidian distance was used) and a region (ellipse) $G^k(\gamma^k(\vx))$.
Thus, we can set the penalty function for feasibility as

\bqa
p(\vx) = \min_{k = 1\ldots m} \Psi\Phi(Z_\alpha+\sqrt{H}(\tilde{\alpha} - \gamma^k(\vx))),
\eqa
where $\Psi, \alpha, \tilde{\alpha}$ and $H$ are running parameters, and $\Phi$ is the cumulative standard Normal distribution.

\begin{figure} 
\centering
  \includegraphics[width=.3\linewidth]{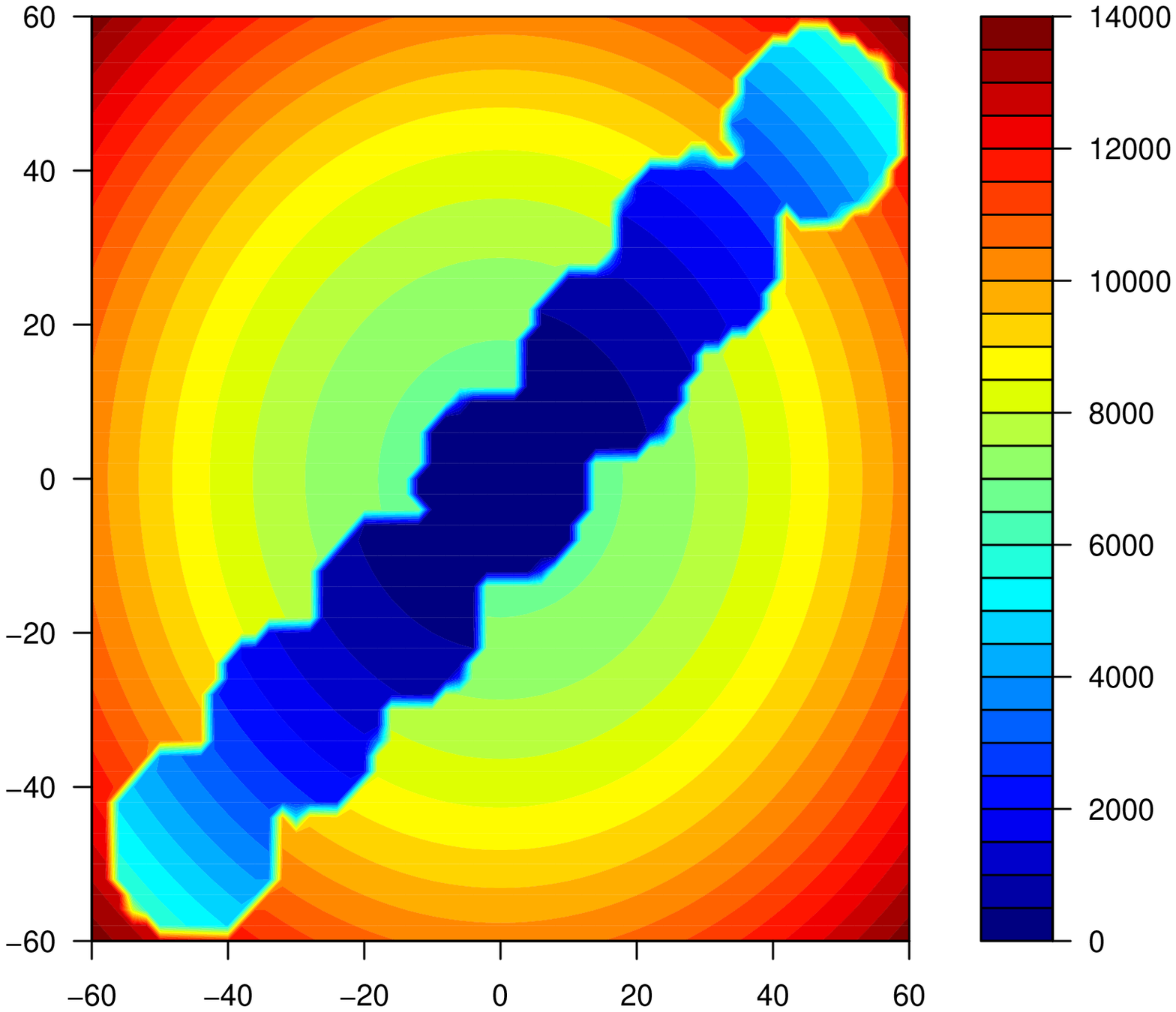}
  \includegraphics[width=.317\linewidth]{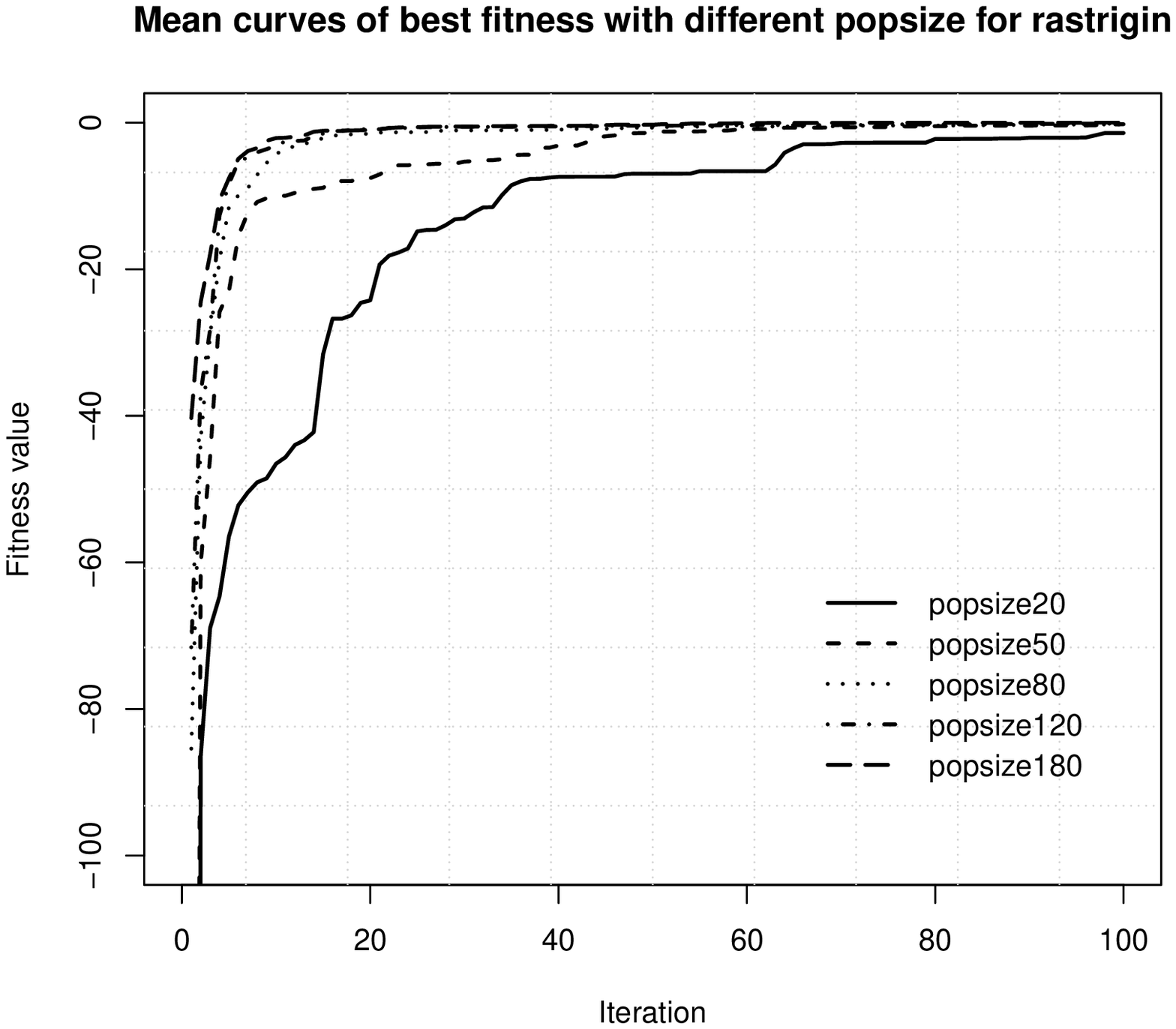}
  \includegraphics[width=.3\linewidth]{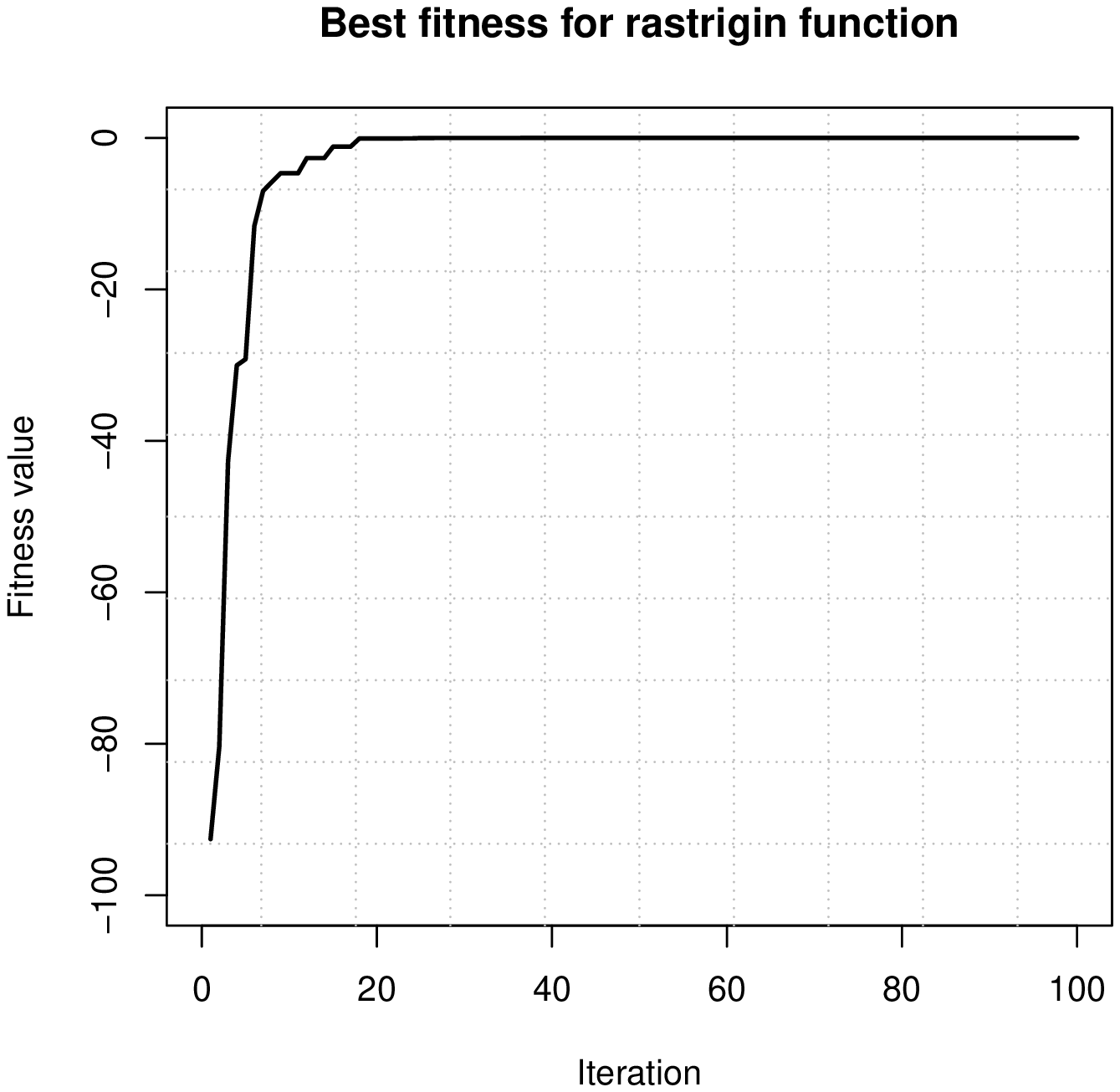}
  \caption{Contour plot, mean fitness curves for several population sizes and mean fitness curve for the ga with population size 80.}\label{fig.Circles}
\end{figure}

In the simulations, we set $\Psi = 7200, \alpha = \tilde{\alpha} = 0.05, H = 10000$. $\Psi = 7200$ was chosen because the maximum of the rastrigin function in the selected search space is $f(60,60) = 7200$, while $H = 10000$ is a reasonably steep slope for the transition to the penalty. Using the function "ga" from the package GA" in "R" (www.r-project.org), we tested several crossover and mutation probabilities, all of which gave similar results, so that in what follows we present the solutions for crossover probability 0.5 and mutation 0.025 since they gave slightly better results. For the stochasticity,  we generated $n = 10$ random observations from each circle center, and obtained the results on shown in Figure \ref{fig.Circles}, where fitness is defined as $-f(\vx)$ (since "ga" runs maximization).

It is clear from the contour plot that the feasible region has much smaller fitness when compared to the rest of the search area. Moreover, the penalty function has facilitated moving from one feasible region to the other, adjusting for stochasticity on the borders. Each of the mean curves in the middle plot  of Figure \ref{fig.Circles} was computed averaging 10 runs of the genetic algorithm using population sizes 20, 50, 80, 120 and 180. The larger the population size is, the faster the algorithm approaches the minimum fitness, however, with a population size of as low as 80 individuals, the algorithm has good performance. 
It can be seen from the plot on the right hand side of Figure \ref{fig.Circles}, where we used a population size of 80 individuals, that the genetic algorithm converges fast towards the target, having a fitness equal to -8.571568e-08 at the end of 100 iterations, close enough to the real solution 0.

\subsection{Regression: Dividing the Plane}

\begin{figure}
\centering
\begin{minipage}{.5\textwidth}
  \centering
  \includegraphics[width=.7\linewidth]{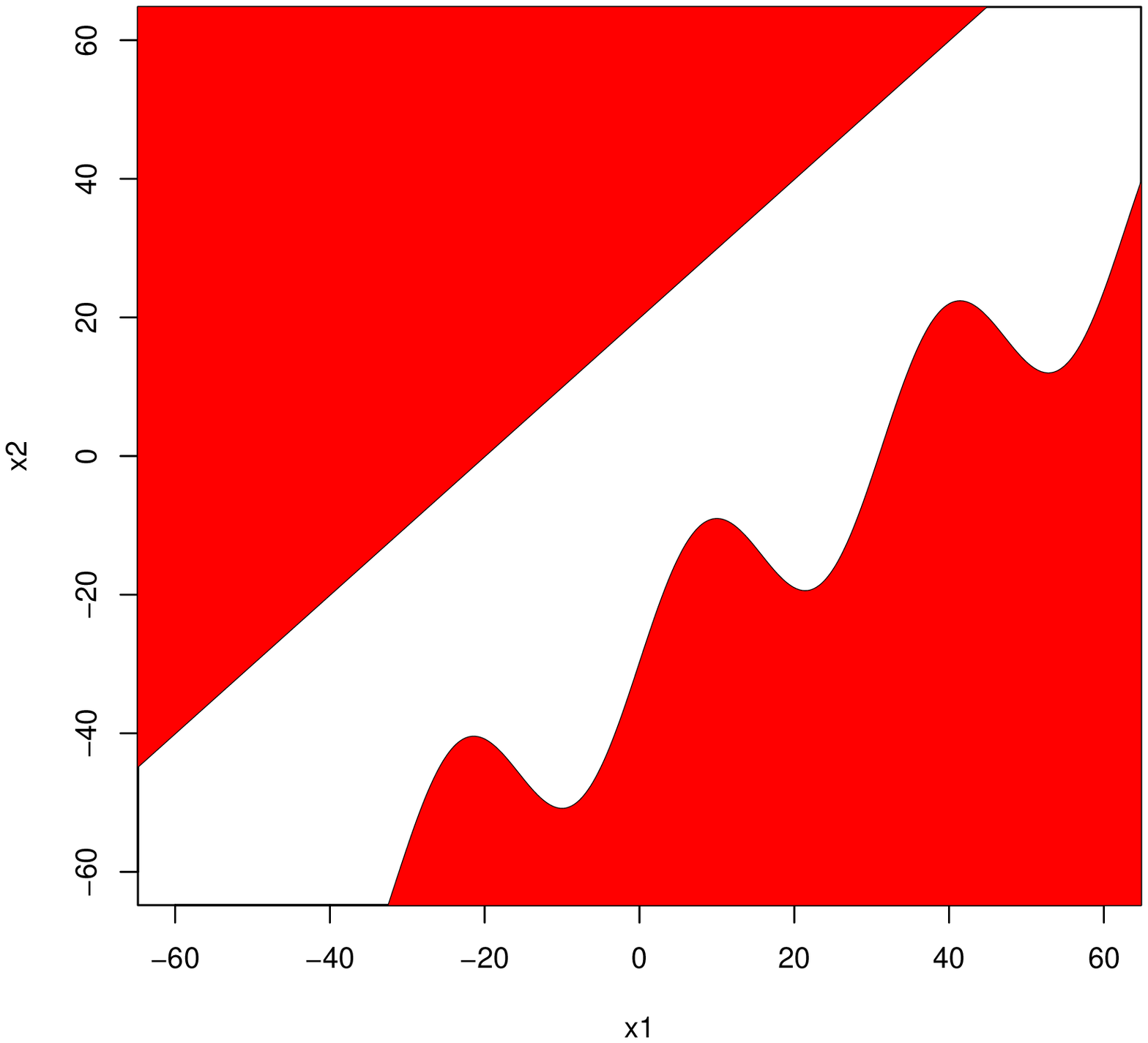}
  \caption{Left: Feasible region in red. Right: Example of observations of $\vX^1$ and $\vX^2$}{}
  \label{Fig.Schwefel_F}
\end{minipage}%
\begin{minipage}{.5\textwidth}
  \centering
  \includegraphics[width=.7\linewidth]{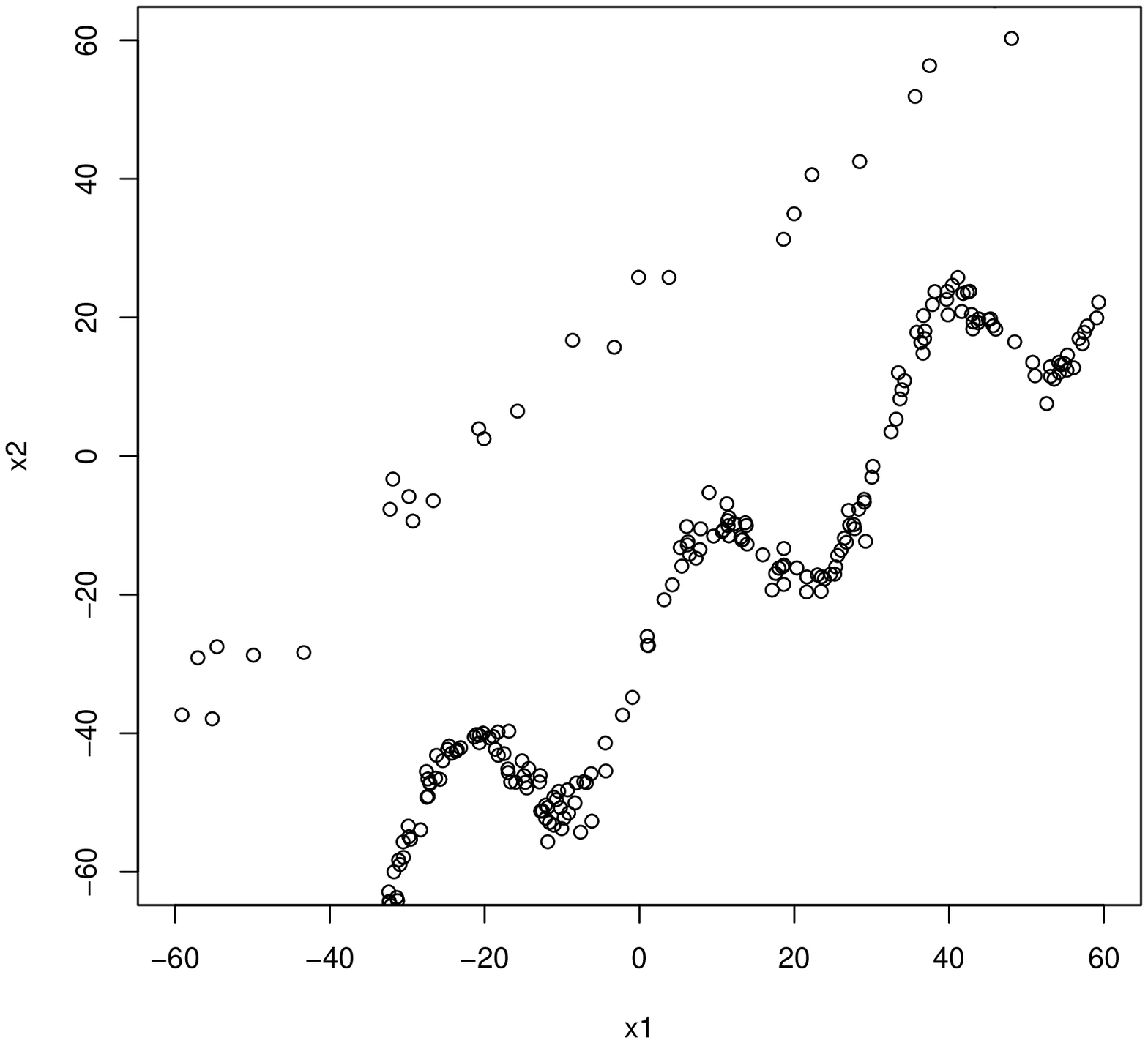}
\end{minipage}
\end{figure}

\indent \indent Suppose that we want to minimize the two-dimensional Schwefel function 
\bqa
f(\vx) = \sum_{k = 1}^{2}\left(\sum_{j = 1}^{k}x_j\right)^2,
\eqa
within the search space $S = \{\vx: -60 < x_1 < 60, -60 < x_2 < 60\}$, however the feasible set is defined by the restrictions
\bqa
&&g_1(\vx) = x_1 - x_2 + 20 \leq 0\\
&&or\\
&&g_2(\vx) = x_1 - 30 + 12\sin(x_1/5) - x_2 \geq 0.
\eqa
Define $F_1$ to be the set of values that satisfy $F_1 =\{\vx: g_1(\vx) \leq 0\}$ and $F_2 = \{\vx: g_2(\vx) \geq 0\}$, so that the feasible set $F = F_1\cup F_2$ (see left plot of Figure \ref{Fig.Schwefel_F}).

Assume that the functions $g_1(\vx)$ and $g_2(\vx)$ are not known, however we have a sample of $n_1 = 30$ observations of $\vX^1 = (X^{11},X^{12})$ and $n_2 = 100$ observations of $\vX^2 = (X^{21},X^{22})$, where $X^{11} \sim U(-60,60)$, $X^{21} \sim U(-60,60)$ and
\bqan\label{equation_reg}
X^{12}_i &=& X^{11}_i + 20 + \epsilon^1_i, i = 1,\ldots,n_1 \nonumber\\
X^{22}_i &=& X^{21}_i - 30 + 12\sin(X^{21}_i/5) + \epsilon^2_i, i = 1,\ldots,n_2,
\eqan
for $\epsilon^1 \sim N(0,5)$ and $\epsilon^2 \sim N(0,2)$. An example of such observations is found in the right plot of Figure \ref{Fig.Schwefel_F}. The equations in (\ref{equation_reg}) naturally characterise a linear and a nonlinear model. In statistical literature, the methodology commonly adopted to estimate these functions is the widely used regression analysis.
If nothing is known about the type of function to be estimated, it is always advisable to use a nonparametric estimate, however we will assume that the the form of $g_1$ is known up to the parameters, but not the form of $g_2$. In this way, $g_1$ will be estimated with a simple linear regression while $g_2$ will be estimated with the Nadaraya-Watson kernel regression. From the linear regression we obtain $\hat{\beta}_0$ and $\hat{\beta}_1$, and from the nonparametric regression we obtain $\hat{g}_2(x_1)$.

In this example, we define the probability $\gamma ^k(\vx)$ that a possible solution point $\vx$ is feasible by
\begin{enumerate}
\item if $\hat{\beta}_0 + \hat{\beta}_1x_1 \leq x_2$, then the probability that $\vx$ is in feasible for the first constraint is approximately equal to $\gamma^1(\vx) \approx 1$
\item if $\hat{\beta}_0 + \hat{\beta}_1x_1 > x_2$, then the probability that $\vx$ is feasible for the first constraint is approximately equal to $\gamma^1(\vx) = 2(1-\Gamma(t,(n_1-2)))$, where $\Gamma(t,(n_1-2))$ is the accumulated t-Student distribution with $n_1-2$ degrees of freedom at point $t$, and \\
$t = (\hat{\beta}_0 + \hat{\beta}_1 x_1 - x_2)/\sqrt{\hat{\sigma}^2(1/n_1 + (x_1 - \bar{X})^2/\sum(X_i - \bar{X})^2)}$.

\item if $x_2 - \hat{m}(x_1) \leq 0$, then the probability that $\vx$ is feasible for the second constraint is approximately equal to $\gamma^2(\vx) \approx 1$
\item if $x_2 - \hat{m}(x_1) > 0$, then the probability that $\vx$ is feasible for the second constraint is approximately equal to $\gamma^2(\vx) = 2(1-\Phi(z))$, where $\Phi$ is the accumulated Gaussian distribution, and $z = (x_2 - \hat{m}(x_1))/\sqrt{||K||_2\hat{\sigma}^2(x_1)/nh\hat{f}_h(x_1)}$.
\end{enumerate}

\begin{figure}[ht] 
\includegraphics[width=.3\textwidth]{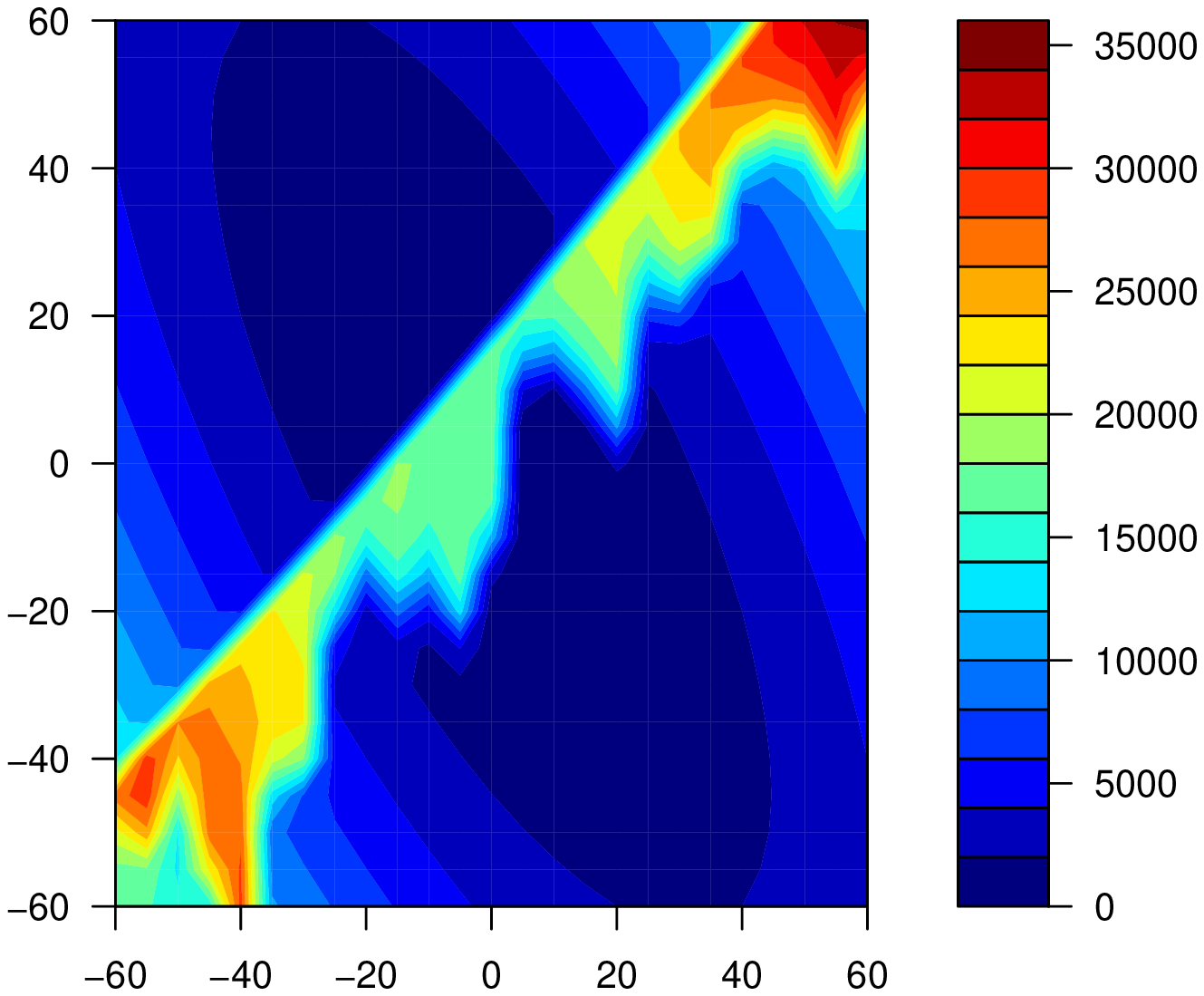}
\includegraphics[width=.317\textwidth]{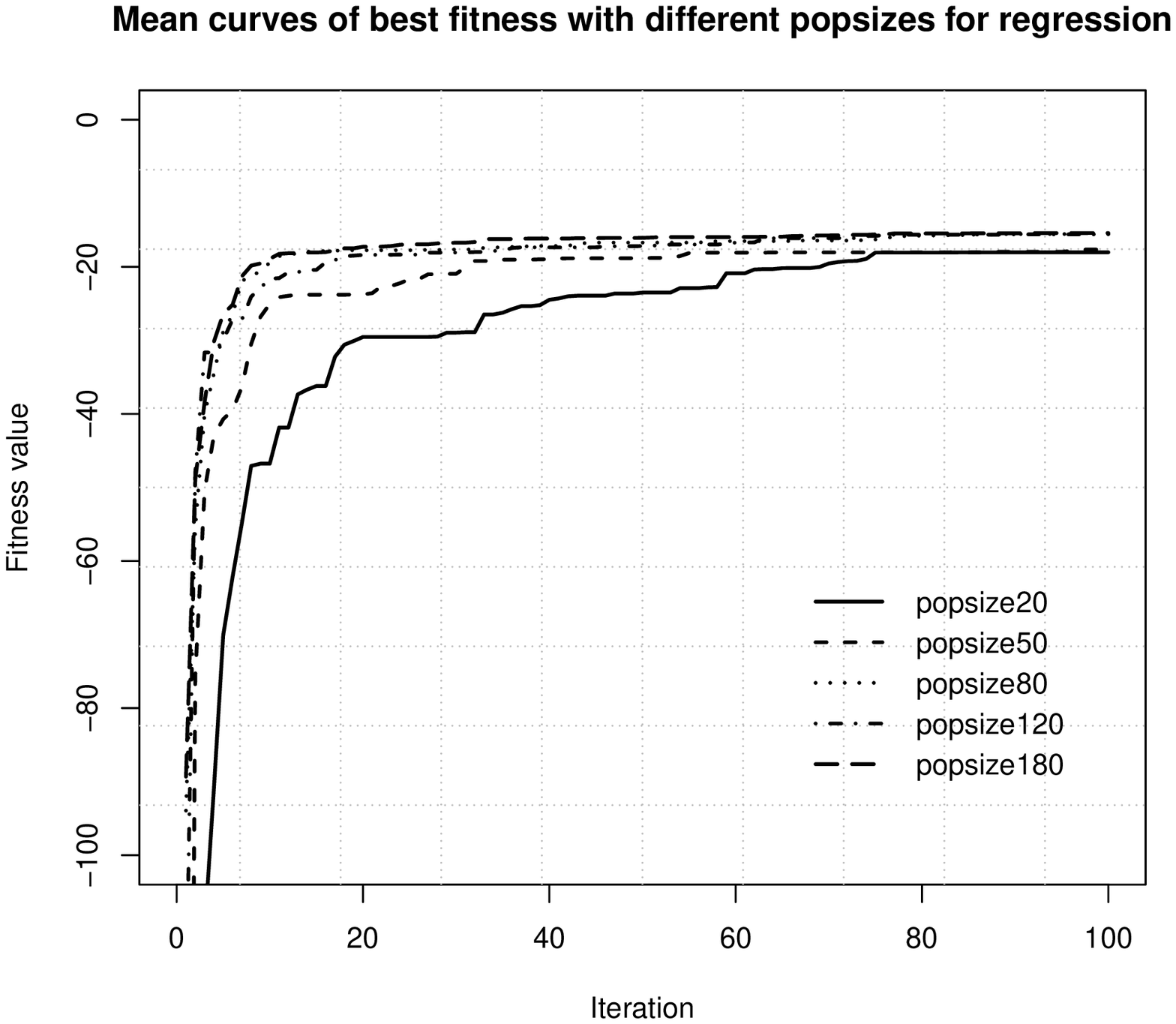}
\includegraphics[width=.295\textwidth]{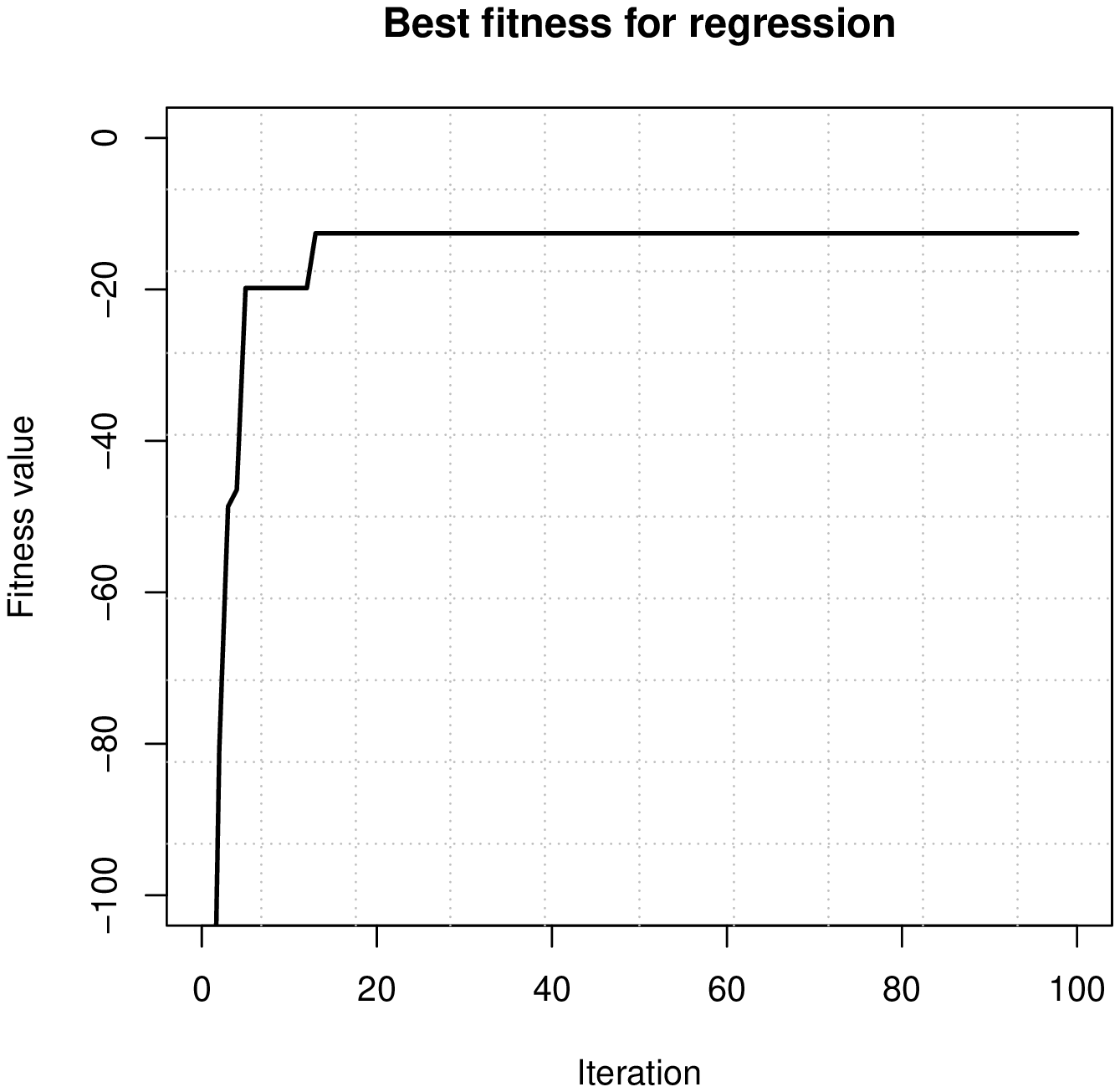}
\caption{Contour plot, mean fitness curves for several population sizes and mean fitness curve for the ga with population size 50.}
\label{fig.DivPlane}
\end{figure}

In the simulations, we set $\Psi = 18000, \alpha = 0.05, \tilde{\alpha} = 0.30, H = 200$. $\Psi = 18000$ was chosen because the maximum of the Schwefel function in the selected search space is $f(60,60) = 18000$. Using the function "ga" in R, again different mutation and crossover probabilities were tested giving similar results, so that we show the results with crossover probability of 0.9 and mutation probability of 0.075.  We generate $n_1 = 30$ observations from the linear regression and $n_2 = 100$ observations for the nonlinear regression, obtaining the results displayed in Figure \ref{fig.DivPlane}.

After 100 iterations, the final solution of the algorithm was the coordinates (3.541831, -3.745075) with fitness -12.58606. It can be seen from the contour plot on Figure \ref{fig.DivPlane}, that this solution is, in the feasible region, as close as possible to the point (0,0) (the unconstrained solution). A population size of as low as 50 gives fast convergence of the algorithm, as can be seen in the mean regression curves in the middle plot of Figure \ref{fig.DivPlane}. The plot on the right hand side of Figure \ref{fig.DivPlane} shows the final result of the genetic algorithm when using a population size of 50 and crossover and mutation probabilities 0.9 and 0.075 respectively. Note that the fitness of the best individual of the population converges to the solution with just under 20 iterations. 


\section{Vehicle Path Planning} \label{sec.robot}
\indent \indent Metaheuristic search methods have recently become popular in applications such as robot or vehicle path planning. The main goal is to find a path with minimum length from a starting to an ending point avoiding obstacles on the way. Many authors have addressed this problem with different strategies, see \citeasnoun{HermanuEtAl2004},  \citeasnoun{Candido},  \citeasnoun{JiaVagners2004}, \citeasnoun{CastilloTrujillo2006},  \citeasnoun{RyersonZhang2007}, \citeasnoun{Xiao-fengEtAl2012}, \citeasnoun{KokEtAl2013} just to cite a few. For example, \citeasnoun{Candido} describes the fitness function as a sum of the lengths of partial paths and proposes a penalty function that penalizes the individuals by counting the number of collisions between the path and the obstacles. He also stores the path as a collection of angles in waypoints, what allows the path to have some abrupt changes. \citeasnoun{HermanuEtAl2004} describe a genetic algorithm which searches for a minimal path going through grids. The computational cost of this approach increases as the size of the cells decreases and the number of allowed headings increases, and moreover, it also allows the vehicle to have abrupt changes on the way. \citeasnoun{JiaVagners2004} propose an algorithm that runs independently with N+1 parallel populations, and at some point, it updates the main population with the best individuals of the other N populations. Even though the probability that the algorithm will be trapped in a local minimum is smaller for this kind of algorithm, they need many more iterations to find the global minimum when considering all parallel populations.

In this paper, we consider only smooth functions to describe the vehicle's trajectory, that is, no abrupt changes in direction are allowed. We adopt the stochastic model studied in \citeasnoun{DiasEtAl2010}, where the sensor of the vehicle observes obstacles with error measurement and a nonparametric method is used to find the best trajectory completely based on data learning. 
In most non-evolutionary search methods, obtaining the exact solution may depend drastically on the initial values. For this reason, we propose using a genetic algorithm to find the parameters that minimize the function corresponding to the minimal path. In section \ref{sec.Stoc.Model} we describe the stochastic scenario and the function that needs to be minimized.

\subsection{The Stochastic Model} \label{sec.Stoc.Model}
\indent \indent As in \citeasnoun{DiasEtAl2010}, we will consider the search for a path with minimum length, which can be represented by the graph of a smooth function $f$ joining the starting and ending point. Without loss of generality, consider the starting point $A = (0, 0)$ and the ending point $B = (b, 0)$, and that we observe $m$ points in ${\mathbb R}^2$
with coordinates $\xi_{\ell} = (w_\ell,v_\ell), l = 1,...,m$. Denote the center of the obstacles by $N = (\xi_1,...,\xi_m)$ and the observed obstacles by $\eta = N + \boldsymbol{\varepsilon}$, where $\boldsymbol{\varepsilon} = (\varepsilon_1, \ldots, \varepsilon_m)$ are the measurement errors. Specifically, we will assume that the region representing the obstacles is
$$F = \cup_{k=1}^{m}B^k((w_k,v_k),r)$$
for some radius $r$, where $B^k((w_k,v_k),r) = \{z \in {\mathbb R}^2; d(z,(w_k,v_k)) < r\}$ and $d$ is the Euclidean distance, but the observations are $(W_k, V_k) = (w_k,v_k) + (\varepsilon_{k  1},\varepsilon_{k 2}), \quad k = 1,...,m$ with
$(\varepsilon_{k 1},\varepsilon_{k 2}) \sim N_2((0,0),  \mathbf{\Sigma}_k)$, $k = 1, \ldots, m$ independent random variables with covariance matrix $\mathbf{\Sigma}_k$ (which can depend on the obstacle ($w_k, v_k$)) given by
\begin{equation}
\mathbf{\Sigma}_k = \left( \begin{array}{cc}
\sigma_{k,1}^{2} & \rho \sigma_{k,1}\sigma_{k,2} \\
\rho \sigma_{k,1}\sigma_{k,2} & \sigma_{k,2}^{2} \\
\end{array} \right).
\end{equation}
This definition of the covariance matrix incorporates several practical situations, such as larger variance for darker spots, increasing variance depending on the distance to the obstacle/threat zone, sensor uncalibrated, etc.

\indent Moreover, we have for each point, $n$ independent readings. Thus, our data is 
composed of $n$ readings of the point process $\eta_i = \{(W_{1,i},V_{1,i}), \ldots, (W_{m,i},V_{m,i})\}$ for $i = 1,\ldots,
n$. Denote $\mathbf{W}_k = (W_{k,1}, \ldots, W_{k,n})$ and $\mathbf{V}_k =
(V_{k,1}, \ldots, V_{k,n})$, $k=1,\ldots,m$.\\
\indent The goal is to find a smooth function $f$ joining $A$ and $B$, avoiding $F$ and with miminum length, which can be represented by
\begin{equation}
\label{eq:v1}
f_{\boldsymbol{\theta}}(x) \,=\, \sum_{j=1}^L \theta_j B_j(x)
\end{equation}
where $B_j$ are cubic B-splines and $\boldsymbol{\theta}$ is a vector of unknown coefficients for a fixed sequence of knots $\boldsymbol{t} = (t_1,...,t_L)$. We consider only functions $f$ belonging to a subset of the Sobolev space $\mathcal{H}^2_2 := \left\{f:[x_a, x_b] \rightarrow \mathbb{R}, \int f^2 + \int (f')^2 + \int (f'')^2 < \infty, f(x_a) = y_a, f(x_b) = y_b\right\}$. This Sobolev space is convenient because it consists of smooth functions that can be well approximated by uniform B-splines.

\indent The proposed estimator for $f_{\boldsymbol{\theta}}$ is the function
\begin{equation}
\label{eq:v1}
f_{\hat{\boldsymbol{\theta}}}^{\gamma}(x) \,=\, \sum_{j=1}^L \hat{\theta_j} B_j(x)
\end{equation}
where $\hat{\boldsymbol{\theta}}$ is the solution of the problem
\begin{equation}
\hat{\boldsymbol{\theta}} = \mbox{arg} \min Q_{\alpha, \psi, r, H, n}(\boldsymbol{\theta}) \label{eq:argmin_stoch}
\end{equation} 
subject to $f_{\boldsymbol{\theta}}^{\gamma}(0) = 0$, $f_{\boldsymbol{\theta}}^{\gamma}(b) = 0$, where
\begin{eqnarray}
\label{eq:stoc}
Q_{\alpha, \psi, r, H, n}(\boldsymbol{\theta}) &=& \int_{0}^{b} \left(1 + (\sum_{j=1}^L
    \theta_j B_j'(t))^{2} \right)^{1/2} dt \\
 && \quad \quad + \psi \,\Phi\left(Z_{\alpha} +
    \sqrt{H}(r-d(\sum_{j=1}^L \theta_j B_j
    (\cdot),F_n^{\gamma})\right). \nonumber
\end{eqnarray}
where $d(f,F) = inf\{d(y,z): y \in \mbox{Graph}(f), z \in F\}$ for $\mbox{Graph}(f) = \{(x , y ): x \in [x_a , x_b ] \mbox{ and } y = f (x)\}$.

\indent The first summand is the length of the function and the second is the penalty for being close to the obstacles. The set $F_n^{\gamma}$ is defined by
\begin{equation}
\label{eq:fat}
F_n^{\gamma} = \bigcup_{k=1}^m G^k_{\gamma^k}(\mathbf{W}_k,\mathbf{V}_k)
\end{equation}
where for each $k=1, \ldots, m$, $G^k_{\gamma^k}(\mathbf{W}_k,\mathbf{V}_k)$ is a $100(1-\gamma)$\% {\it confidence ellipse} based on $n$ readings for the $k$th point $(\mathbf{W}_k,\mathbf{V}_k)$, defined as the ellipse formed by the points $(x,y)$ which satisfy the equation
$$n((\overline{\mathbf{W}}_k,\overline{\mathbf{V}}_k) - (x,y))^t\mathbf{\Sigma}_k^{-1}((\overline{\mathbf{W}}_k,\overline{\mathbf{V}}_k) - (x,y))
\le \chi^2_2(\gamma).$$

Suppose that besides the constraints of the obstacles in the way, the autonomous vehicle has to follow a pre-determined path, from which it can not deviate. Using the union of sets definition of Section \ref{sec.union}, we can add the union of $t$ restrictions that correspond to the constraints of the corridor (path). Hence we let
\bqa
F_n^{\gamma} = \bigcup_{k=1}^m G^k_{\gamma^k}(\mathbf{W}_k,\mathbf{V}_k)\bigcup_{\ell=1}^t H^\ell_{\gamma^\ell}(\vX_\ell, \vY_\ell),
\eqa
where $\vX_\ell, \vY_\ell, \ell = 1, \ldots, t$, are random variables that can be used to estimate the path, and $H^\ell_{\gamma^\ell}$ define the form of the restrictions. The penalty function can then be modified to incorporate this new set of restrictions. 

Note that each individual of the population of the genetic algorithm is represented by $\boldsymbol{\theta}$, and that the feasibility region is composed by the union of regions on the search space $\boldsymbol{\Theta}$ such that, the function $f_{\boldsymbol{\theta}}$ is distant (with probability $\gamma$) from the obstacles. The union of balls corresponding to the obstacles $F =  \cup_{k=1}^{m}B^k((w_k,v_k),r)$, or its stochastic version $F_n^{\gamma}$, does not correspond directly to the feasible region, for it is not defined explicitly with the vector $\boldsymbol{\theta}$, but it can be translated into regions in the search space $\boldsymbol{\Theta}$ with respect to $f_{\boldsymbol{\theta}}$. Also, the penalty function in (\ref{eq:stoc}) can be modified to be in the form of penalties defined in (\ref{penalty}), just by noting that $d(\sum_{j=1}^L \theta_j B_j    (\cdot),F_n^{\gamma})$ is the minimum distance between the graph of $f$ and any point in $F_n^{\gamma}$.

Hence, this is a constrained optimization problem that can be solved using a genetic algorithm. To be specific, the algorithm searches for the minimum of $Q_{\alpha, \psi, r, H, n}(\boldsymbol{\theta})$, the length of the estimated trajectory penalized according to the degree of violation of the probability that the solution is feasible. 

\subsection{Numerical Examples} \label{sec.Gen.Alg.Robot}
\indent \indent We simulate obstacle fields with obstacles scattered on the path the vehicle must travel. In all simulations, the goal is to go from point A to point B, taken in all the examples to be (0,0) and (150, 0) respectively. We assume, without loss of generality, that the obstacles are circles of radius 4. The radar observations of the center of the obstacles were generated from a bivariate Gaussian distribution and for illustration purposes, we exhibit only the cases where we have 10 and 45 readings for each obstacle. 
In all the plots, black circles represent the real obstacles, the red crosses are the mean of the observed obstacles, and the confidence ellipses are drawn in grey. Each trajectory was estimated as in (\ref{eq:v1}) using 3 B-splines basis functions, restricting the estimated trajectory to a very smooth curve and reducing computational cost. Nevertheless, the degree of smoothness for a solution can be adapted to a specific problem by choosing the number of basis functions adequately. The tuning parameters for the penalty function are taken to be $\gamma = 0.05, \psi = 50, \alpha = 0.01, H = 200$. The results presented in this section are based on 100 generations of the genetic algorithm with stochastic feasibility as described in Section \ref{sec.robot}, using the "ga" function for a real-coded algorithm from package GA in "R" (www.r-project.org).

\begin{figure}[ht!]
     \begin{center}
        \subfigure[]{%
            \label{fig:first}
            \includegraphics[width=0.4\textwidth]{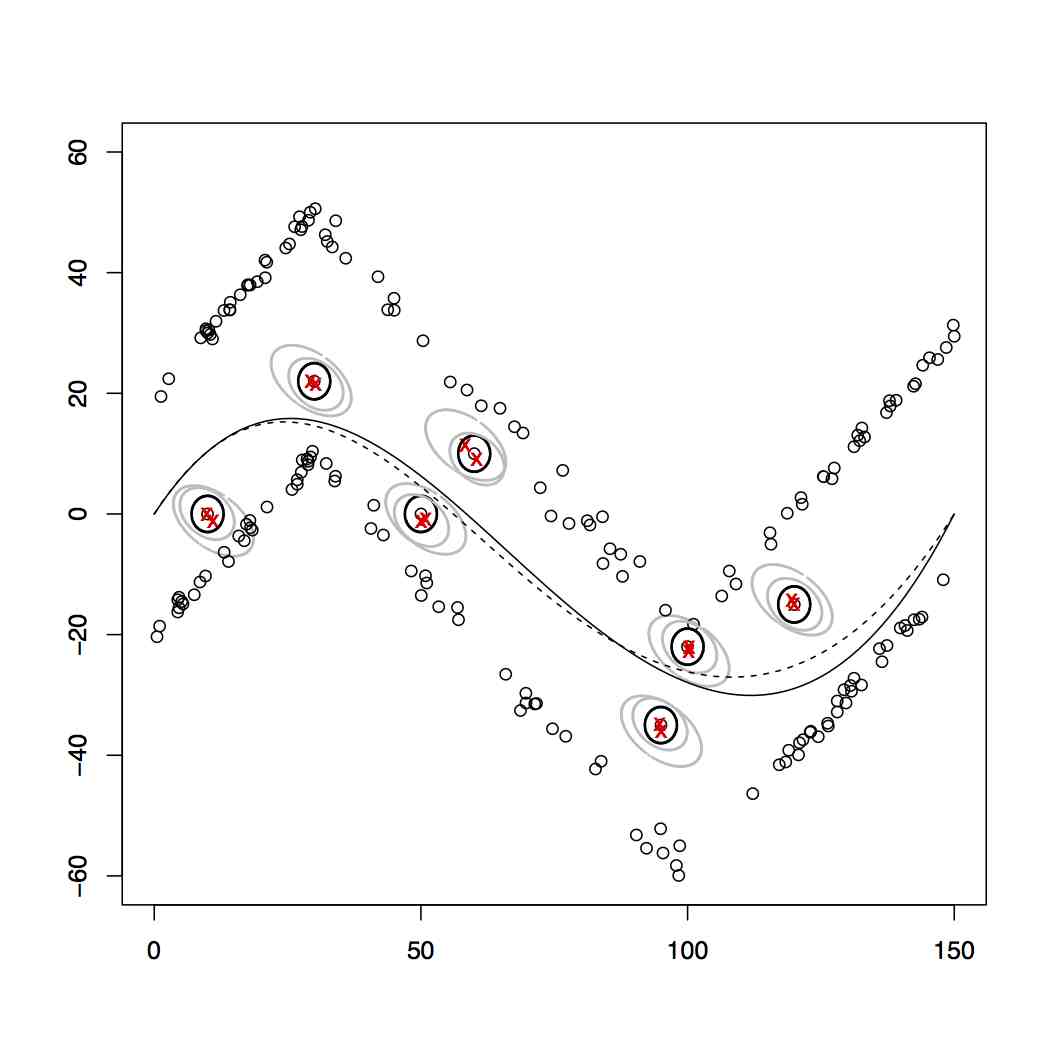}
        }%
        \subfigure[]{%
            \label{fig:fourth}
            \includegraphics[width=0.4\textwidth]{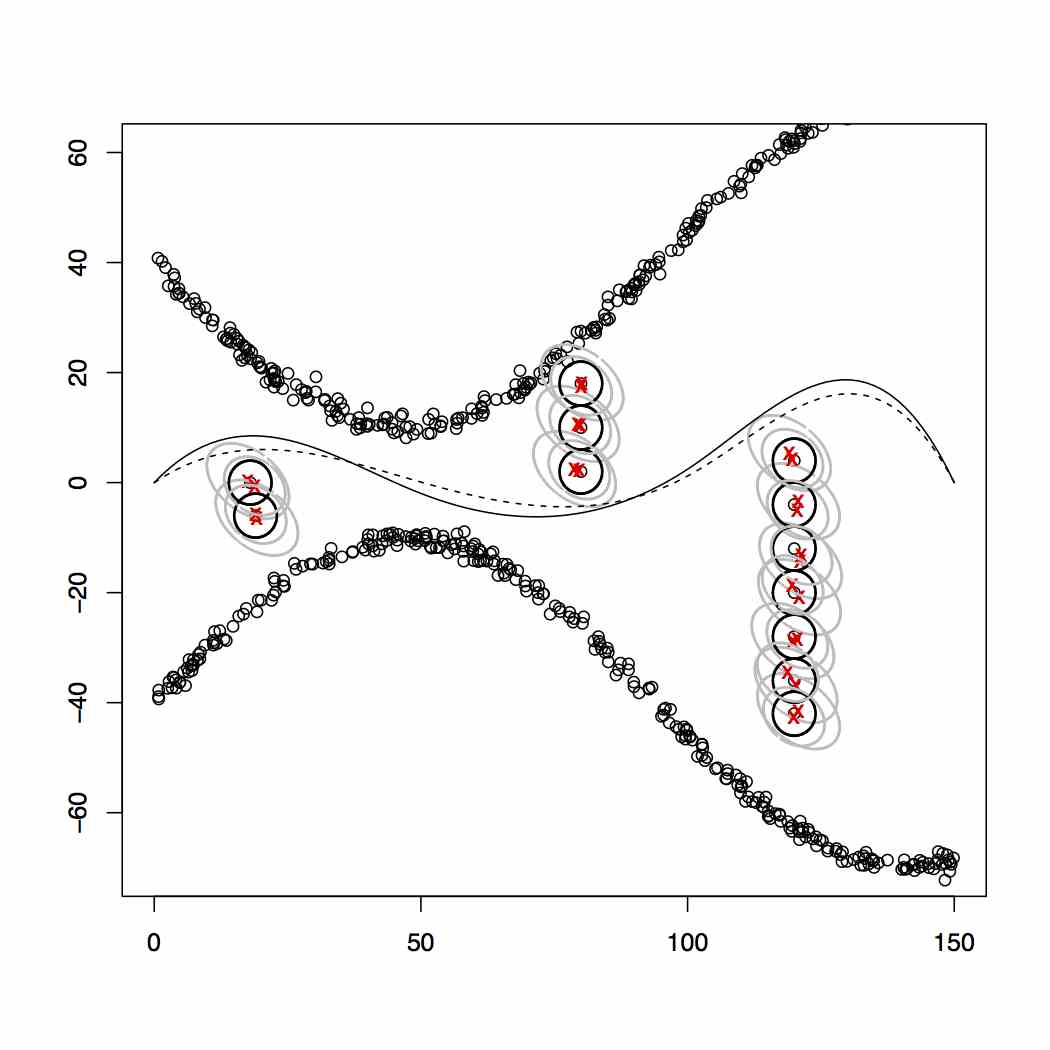}
        }%
    \end{center}
    \caption{%
        Estimated trajectories after 100 generations based on 10 readings (solid curve) and 45 readings (dashed line) of the obstacles. The obstacles are normally (Gaussian) distributed with (a) $\rho = -0.8$, $\sigma_{k,1} = 4$ and $\sigma_{k,2} = 6$; (b)  $\rho = -0.8$, $\sigma_{k,1} = 4$ and $\sigma_{k,2} = 5$.
     }%
   \label{fig:subfigures}
\end{figure}

In the first scenario, we simulate the corridor limits in the following way: 30 independent observations of each of the random variables $(X_1,\ldots, X_6)$ are randomly generated from uniform distributions (denoted by $U(.,.)$) such that 
\bqa
X_1, X_2 &\sim& U(0, 30),\\
X_3, X_4 &\sim& U(30, 100),\\
X_5, X_6 &\sim& U(100, 150).
\eqa
Using these observations, we generate the random variables $(Y_1, \ldots, Y_6)$ from the following linear regressions
\bqa
Y_1 &=& 20 + X_1 + \epsilon_1,\\
Y_2 &=& -20 + X_2 + \epsilon_2,\\
Y_3 &=& 80 - X_3 + \epsilon_3,\\
Y_4 &=& 40 - X_4 + \epsilon_4,\\
Y_5 &=& -120 + X_5 + \epsilon_5,\\
Y_6 &=& -160 + X_6 + \epsilon_6,
\eqa
where $\epsilon_1, \ldots, \epsilon_6$ are independent measurement errors, which in this example have standard Gaussian distributions, except for $\epsilon_3, \epsilon_4$ which are Gaussian but with mean 0 and standard deviation 2. The estimated linear regressions are used to build each $H^\ell_{\gamma^\ell}(\vX_\ell, \vY_\ell), \ell = 1, \ldots, 6$, restricting the vehicle to the path. 
The results are displayed in graph (a) of Figure \ref{fig:subfigures}.

For the remaining simulations, we estimate the corridor limits with nonparametric regressions using local quadratic fits. Hence, the algorithm learns the directions from the data, without any assumptions about the form of the corridors. Also, we generate 300 of the random variables
\bqa
X_1, X_2 &\sim& U(0, 150),
\eqa
to be used in constructing the pathways of the remaining numerical examples.

In the second scenario, we use the observations from $(X_1, X_2)$ to generate the random variables $(Y_1, Y_2)$ from the following nonlinear relations
\bqa
Y_1 &=& 20 + 30\sin(X_1/45) + \epsilon\\
Y_2 &=& -20 + 30\sin(X_1/45) + \epsilon,
\eqa
where $\epsilon$ the measurement error, whose distribution is standard Gaussian. These two nonlinear regressions compose the corridor limits where the vehicle can not deviate, i.e,\\ $H^\ell_{\gamma^\ell}(\vX_\ell, \vY_\ell), \ell = 1, 2$. The results are shown in graph (b) of Figure \ref{fig:subfigures}.
The third scenario has the following nonlinear regressions defining the pathway
\bqa
Y_1 &=& 1 + 30\cos(X_1/25) + \epsilon\\
Y_2 &=& -40 + 30\cos(X_2/25) + \epsilon,
\eqa
where $\epsilon$ is standard Gaussian. The results are in graph (c) of Figure \ref{fig:subfigures}.
In the last graph of Figure \ref{fig:subfigures}, the limits of the corridors are defined by the equations
\bqa
Y_1 &=& 40 - 30\sin(X_1/30) + \epsilon\\
Y_2 &=& -40 + 30\sin(X_2/30) + \epsilon.
\eqa

All the plots in Figure \ref{fig:subfigures} show that in the presence of random variability, the paths tend to stay further from the obstacles, increasing the length of the trajectory. However, as the number of readings increases, the algorithm can compute smaller confidence regions from data learning, establishing shorter trajectories.

\section{Conclusion}
\indent \indent In most practical situations, observed data have measurement errors. In this paper, we presented an approach for genetic algorithms to solve constrained optimization problems in the presence of stochastic  feasibility regions. The proposed method allows the algorithm to learn from the data by building confidence regions of feasibility that the solutions should be restrained to. As the number of observations grows, the confidence regions shrink to the true feasibility regions, defining an automatic data learning algorithm.
We proposed a smooth penalty function that penalizes individuals by the probability of violation of the confidence regions. The probability of feasibility is problem specific and can be computed by characterizing the confidence level of the region that intersects with each possible solution of the search space. 
We also proposed working with a slightly different definition of the constrained optimization problem. In this definition, we write the feasibility regions as the union of disjoint sets, facilitating mathematical notation.
Lastly, we applied the genetic algorithm with stochastic feasibility regions to the problem of trajectory planning for unmanned vehicles. The algorithm searches for the trajectory with minimum length, while avoiding obstacles on the way and keeping within pathway limits, which are observed with measurement errors.

{\bf Acknowledgements} This paper was partially supported by FAEPEX Grant 519.292 (573/13), CNPq 302182/2010-1, Fapesp: 2013/07375-0, Fapesp: 2013/00506-1 and CAPES.
\bibliography{mybib}  

\end{document}